\documentclass[twocolumn,noshowpacs,amsmath,amssymb]{revtex4}

\usepackage{graphicx}
\usepackage{graphics}
\usepackage{amsmath}
\usepackage{amstext}
\usepackage{enumerate}
\usepackage{natbib}

\begin{document}


\title{Generation of neutral atomic beams \\ utilizing  photodetachment by high power diode laser stacks} 
\author{A. P. O'Connor}
\email[]{aodh.oconnor@mpi-hd.mpg.de}
\affiliation{Max-Planck-Institut f\"ur Kernphysik, 69117 Heidelberg, Germany}

\author{H. Bruhns\footnote{Present address: Inficon GmbH, Bonner Str.\,498, 50968 K\"oln, Germany}}
\affiliation{Columbia Astrophysics Laboratory, Columbia University, New York, NY 10027, U.S.A.}

\author{{F. Grussie}}
\affiliation{Max-Planck-Institut f\"ur Kernphysik, 69117 Heidelberg, Germany}

\author{T. P. Koenning}
\affiliation{Dilas Diode Laser Inc., Tucson, AZ 85715, U.S.A}

\author{K. A. Miller}
\affiliation{Columbia Astrophysics Laboratory, Columbia University, New York, NY 10027, U.S.A.}

\author{N. de Ruette\footnote{Present address: Department of Physics, Stockholm University, SE-106 91 Stockholm, Sweden \\ }}
\affiliation{Columbia Astrophysics Laboratory, Columbia University, New York, NY 10027, U.S.A.}

\author{J. St\"utzel}
\affiliation{Columbia Astrophysics Laboratory, Columbia University, New York, NY 10027, U.S.A.}

\author{D. W. Savin}
\affiliation{Columbia Astrophysics Laboratory, Columbia University, New York, NY 10027, U.S.A.}

\author{X. Urbain}
\affiliation{Institute of Condensed Matter and Nanosciences, Universit\'e Catholique de Louvain, Louvain-la-Neuve B-1348, Belgium}

\author{H. Kreckel}
\affiliation{Max-Planck-Institut f\"ur Kernphysik, 69117 Heidelberg, Germany}

\date{\today}

\begin{abstract}
We demonstrate the use of high power diode laser stacks to photodetach fast hydrogen and carbon anions and produce ground term neutral atomic beams. We achieve photodetachment efficiencies of $\sim$7.4\% for H$^-$ at a beam energy of 10\,keV and $\sim$3.7\% for C$^-$ at 28\,keV. The diode laser systems used here operate at 975\,nm and 808\,nm, respectively, and provide high continuous power levels of up to 2\,kW, without the need of additional enhancements like optical cavities. The alignment of the beams is straightforward and operation at constant power levels is very stable, while maintenance is minimal. We present a dedicated photodetachment setup that is suitable to efficiently neutralize the majority of stable negative ions in the periodic table.   
\end{abstract}


\maketitle 

\section{Introduction} \label{sec.introduction}

Interactions of neutral atomic species with other atoms, molecules, and solids are of fundamental interest in many branches of physics research. For example, while none of the natural elements occur in atomic form under normal terrestrial conditions (except for the rare gases and mercury), neutral atoms are routinely observed in space. In interstellar clouds, neutral species like H, D, C, and O are abundant \cite{wooden04} and their reactions with ions and neutrals contribute to the build-up of complex molecules in space and thus remain an active field of research \cite{wakelam09}. Different experimental techniques have been pursued to study ion-neutral \cite{snow08} and neutral-neutral \cite{urbain91, daranlot11} collisions. Often experiments with neutral atoms are carried out in flow tubes, where neutral H and N atoms can be created in microwave discharges, while O atoms are prepared by exchange reactions with nitrogen \cite{snow08}. The neutral species in these measurements are, however, embedded in the gas flow and not freely available for other applications or experiments. 
Experimental data on reactions involving neutral carbon atoms are extremely rare, mainly due to the difficulties of creating neutral C beams (a brief overview can be found in \cite{ocon15}). The method presented here overcomes these difficulties and thus facilitates detailed cross section and rate coefficient measurements with neutral C atoms \cite{ocon15}).

For many technical applications, the generation of well-defined neutral atomic beams is also highly desirable. For example, the use of atomic hydrogen is an established method for substrate cleaning in semiconductor research \cite{hirsch97}.  In nuclear fusion devices, hydrogen or deuterium beams can be used for plasma heating and control, as neutral particles can enter the plasma without being deflected by the strong magnetic fields. For the ITER project, it is foreseen that neutral beams will inject up to 50\,MW of power into the fusion plasma \cite{hemsworth09}. For this application, the creation of intense neutral beams by stripping of H$^-$ or D$^-$ in gas cells has been explored, however, the efficiency of this approach is limited and the pumping needs of large scale gas cells may lead to prohibitive power consumption levels \cite{kovari10}. An alternative path of neutralization would be the use of intense laser light  \cite{fink75}.  Recent design studies for ITER are based on the use of Nd:YAG lasers combined with highly reflective cavities. However, it has been recognized that the technical difficulties in maintaining the beam quality and mirror reflectivities under harsh conditions would be immense \cite{kovari10}. 

Neutralization of negative hydrogen beams by intense laser fields has been demonstrated before \cite{vanzyl76, havener89}, and it has been used extensively at Oak Ridge National Laboratory to study interactions between highly charged ions and atomic hydrogen (see, e.g., \cite{bruhns08, draganic11}). In these experiments the negative hydrogen beam traverses the inner cavity of a Nd:YAG laser, which is used for photodetachment. As the fundamental wavelength of Nd:YAG lasers lies at 1064\,nm, the photon energy is sufficient to photodetach H$^-$ and D$^-$, but it is below the photodetachment threshold for  C$^-$ and O$^-$. 


Here we demonstrate the generation of neutral atomic beams utilizing diode laser stacks to photodetach negative ions. Diode laser stacks are robust and require little maintenance. They provide high power levels without using enhancement methods such as optical cavities. Therefore, some of the more stringent requirements, that the use of optical cavities impose on the operating conditions, can be avoided, and long-term stability is greatly improved.  


\section{The photodetachment process}\label{PD_Process}

\subsection{Basic considerations}

Photodetachment studies have been the main source of information on the electron affinities and the structure of negative ions; and the results have been collected in several reviews \cite{andersen99, andersen04}. While an in-depth discussion of the photodetachment process is beyond the scope of the present publication, here we give a brief overview of those aspects that are relevant for the creation of atomic neutral beams.

The outermost electron of a negative atomic ion is bound to the neutral host atom by polarization forces. These are weak compared to the $1/r$ Coulomb potential that binds the inner electrons. Therefore, most atomic negative ions have only a single bound state, and electron affinities (corresponding to the binding energy of the outer electron) are generally about an order of magnitude lower than the first ionization potential of the host atom. Single-photon electron detachment of a ground state negative ion becomes possible when the energy  of an individual photon $h\nu$ exceeds the electron affinity $E_{\rm EA}$ (here, $h$ is Planck's constant and $\nu$ is the frequency of the photon). At photon energies close to the threshold $E_{\rm th}$, the cross section $\sigma_{\rm PD}$ is usually described by the Wigner threshold law \cite{wigner1948}
\begin{equation}
\sigma_{\rm PD} = (h\nu-E_{th})^{\ell+1/2}\,,
\end{equation}
where $\ell$ is the angular momentum of the detached electron. Angular momentum is conserved during the detachment process and the behavior of the cross section at threshold is dominated by the angular momentum of the departing electron, where $\ell=0$ corresponds to an {\em s}-wave, $\ell=1$ corresponds to a {\em p}-wave, etc. When more than one partial wave is permitted by the selection rules, the lower $\ell$-value typically dominates at threshold. At higher photon energies, detachment into excited states of the neutral atom may become possible, resulting in the opening of additional thresholds.

\begin{table}[b]
\caption{\label{spin}Table of electron affinities ($E_{\rm EA}$), negative ion electronic states, and threshold photodetachment wavelengths ($\lambda_{\rm th}$) of all elements in the first three rows of the periodic table that form stable negative ions \cite{andersen99}. Electron affinities of metastable states are given in parentheses. \label{Tab:EAffinity}}
\begin{ruledtabular}
\begin{tabular}{lclcc}
&$E_{\rm EA}$(eV)&Neg. ion state&$\lambda_{\rm th}$(nm)\\
\hline 
H  & 0.754 & $1s^2\,^1S_0$ & 1643 \\
Li & 0.618 & $2s^2\,^1S_0$ & 2006 \\
B  & 0.290 & $2p^2\,^3P_0$ & 4275 \\
C  & 1.262 & $2p^3\,\,^4S_{3/2}$ & 982 \\
   & (0.033)      & $2p^3\,^2D_{5/2,3/2}$ & 37571 \\
O  & 1.461 & $2p^5\,^2P_{3/2}$ & 848 \\ 
F  & 3.401 & $2p^6\,^1S_0$    & 364\\
Na & 0.548 & $3s^2\,^1S_0$    & 2262 \\
Al & 0.432 & $3p^2\,^3P_0$    & 3484\\
   & (0.109)      & $3p^2\,^1D_2$ &  11375\\
Si & 1.390 & $3p^3\,^4S_{3/2}$ & 892\\
   & (0.527)      & $3p^3\,^2D_{3/2}$ & 2353\\
   & (0.029)      & $3p^3\,^2P_{1/2}$ & 42753\\
P  & 0.747 & $3p^4\,^3P_2$       & 1660\\
S  & 2.077 & $3p^5\,^2P_{3/2}$    & 597\\
Cl & 3.613 & $3p^6\,^1S_0$       & 343\\
\end{tabular}
\end{ruledtabular}
\end{table}

Some negative ions possess excited states, which have a lower electron detachment threshold. However, it is difficult to use these for the production of stable neutral beams, as standard negative ion sources typically yield only a very small fraction of ions in metastable levels (see, e.g., \cite{scheer98} and \cite{takao07} for the case of metastable  C$^-$). 

Table\,\ref{Tab:EAffinity} lists the electron affinities, electronic states, and threshold wavelengths for those elements in the first three rows of the periodic table that form stable negative ions. The electron affinities range from $0-3.6$\,eV, with chlorine being the most electronegative element in the periodic table. It is interesting to compare these values to the wavelength of high power diode laser stacks. In this work, we present data taken with two different laser systems at 975\,nm and 808\,nm, respectively. Cost-effective solutions for high power diode stacks at wavelengths shorter than 800\,nm (corresponding to $h\nu\geq 1.55$\,eV) are currently not commercially available, although promising new materials that can generate laser light at 766\,nm and 793\,nm are under development. Considering all elements in the periodic table up to $Z=86$, most can form stable negative ions and only 12 elements have known electron affinities $>$$1.55$\,eV (the rare earths, with $Z=58-71$, are not well-studied). That means that high power diodes can be used to neutralize the majority of stable negative atomic ions.

\begin{figure}[t]
\begin{center}
\includegraphics[width=0.45\textwidth]{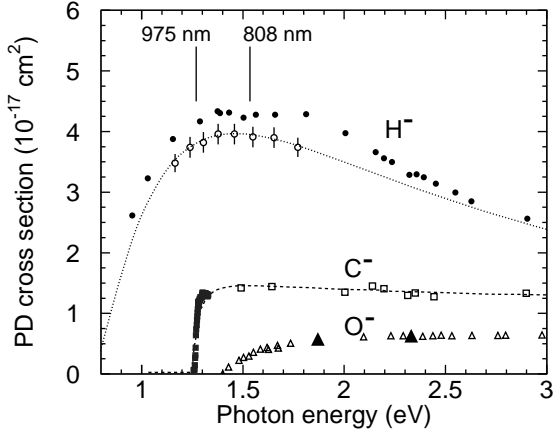} \caption{Measured photodetachment (PD) cross sections for H$^-$ (solid circles \cite{Smith1959} and open circles \cite{gene15}), C$^-$ (squares), and O$^-$ anions (triangles). The dotted line shows the calculated H$^-$ cross section taken from \cite{venuti97}. The solid squares are relative experimental data for C$^-$ \cite{feldmann77} that have been scaled to roughly match the absolute data depicted by the open squares and the threshold fit depicted by the dashed line, which were both taken from \cite{seman1962}. The O$^-$ data shown by open triangles are taken from \cite{branscomb65}, the two large solid triangles depict recent measurements taken from \cite{hlavenka09}. The vertical solid lines represent the laser wavelengths used in the present study.\label{Fig:PD_Efficiencies}}
\end{center}
\end{figure}

The absolute photodetachment cross sections for negative ions of hydrogen, carbon, and oxygen are shown in Fig.\,\ref{Fig:PD_Efficiencies} and compared to the laser wavelengths that were used in the present work. These elements represent three of the four most abundant elements in the universe (the fourth is helium, which forms only metastable anions). Hydrogen, carbon and oxygen are thus of paramount importance for astrophysical studies. The hydrogen cross section shows a very broad maximum, such that any photon energy between 1.2\,eV and 2.0\,eV is suitable.  The cross sections for C$^-$ and O$^-$ show the expected simple threshold behavior and the exact laser wavelength is rather uncritical, as long as the photon energy is above the threshold. 

\subsection{Crossed- and inclined-beams photodetachment}

The photodetachment probability $\eta$ for an anion passing through a laser field depends on the product of the wavelength-dependent cross section $\sigma_{\rm PD}$ and the photon flux $\Phi$, integrated over the interaction time $t$
\begin{equation}\label{Eqn:PD1}
\eta=1-\exp{\int _{-\infty} ^{+\infty}- \sigma_{\mathrm{PD}} \Phi(t) dt}\,.
\end{equation}
Here we consider two simple, illustrative cases in order to derive the relevant parameters determining the photodetachment efficiency for a collimated anion beam.

\begin{figure}[b]
\begin{center}
\includegraphics[width=0.5\textwidth]{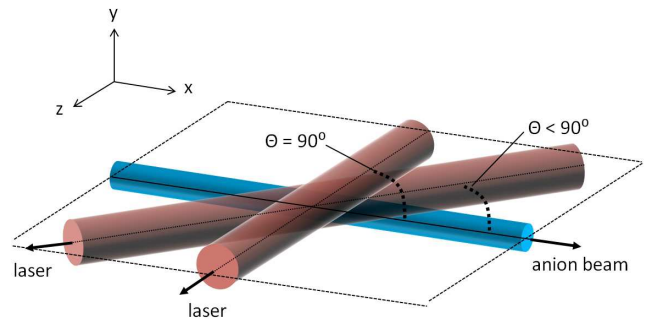} \caption{Photodetachment of an anion beam traveling in $x$-direction with a crossed ($\Theta=90^\circ$) or inclined ($\Theta<90^\circ$) laser beam.\label{Fig:pd-principle}}
\end{center}
\end{figure}

First, let us assume that the ion beam (travelling in the $x$-direction in a Cartesian coordinate system) is crossed at an angle of $90^\circ$ in the $x$-$z$-plane, by a laser beam that is larger in diameter (Fig.\,\ref{Fig:pd-principle}). If the ions travel at constant velocity and both beams are continuous, then the situation is stationary, and instead of integrating over the interaction time, we can use the $x$-coordinate, giving
\begin{equation}\label{Eqn:PD2}
\eta(y,z) = 1- \exp\left[-{\frac{\sigma_{\rm PD}}{v_{\rm ion}}\int_{-\infty}^{+\infty} \Phi(x,y,z)\,dx}\right]\,,
\end{equation}
where $v_{\rm ion}$ is the ion beam velocity. Now we assume that the laser beam propagates in the $z$-direction and has a Gaussian profile with a diameter $2w$ (corresponding to the size of the beam at which the irradiance drops to $1/e^2$ with respect to the value on the propagation axis). The photon flux in this case can be parametrized by
\begin{equation}
\Phi(x,y) = \frac{2\,P}{h\nu\pi w^2} \exp\left[-\frac{2(x^2+y^2)}{w^2}\right]\,,
\end{equation}
where $P$ stands for the laser power. Solving the integral, given by Eq.\,\ref{Eqn:PD2}, yields
\begin{equation}
\eta^{\rm crossed}(y)= 1- \exp{\left[{-\sqrt{\frac{2}{\pi}}\frac{\sigma_{\rm PD}}{v_{\rm ion}}\frac{P}{h\nu w} \exp{\left(-\frac{2y^2}{w^2}\right)}}\right]}\,.
\end{equation}
To simplify matters further, we can assume that we are operating below the saturation regime and use only the leading orders of the series expansion for the first exponential ($e^x=1+x+...$), resulting in
\begin{equation}\label{Eq:pdfinal1}
\eta^{\rm crossed}(y) \approx {\sqrt{\frac{2}{\pi}}\frac{\sigma_{\rm PD}}{v_{\rm ion}}\frac{P}{h\nu w} \exp{\left(-\frac{2y^2}{w^2}\right)}}\,.
\end{equation}
This expression must be averaged over the ion beam profile along the $y$ direction (perpendicular to the intersection plane) in order to obtain the full detachment efficiency. The total efficiency is thus inversely proportional to the laser beam width (and not to the square of it, as one could guess from the corresponding increase of the irradiance), provided that the anion beam is contained within the vertical profile of the laser beam. In order to optimize the global neutralization efficiency for a given ion beam size and velocity, it is advisable to adjust the vertical laser beam size to roughly match the size of the ion beam. If a good match can be achieved, then, other than the laser power and frequency, Eq.\,\ref{Eq:pdfinal1} leaves little room for optimization. However, one has to keep in mind that in our derivation so far we have assumed right angles between both beams.

If we introduce as a second case an angle $\Theta<90^\circ$  (Fig.\,\ref{Fig:pd-principle}), the effective overlap length increases by a factor $1/\sin\Theta$ (assuming that the laser beam profile is constant along the overlap region), resulting in a larger photodetachment efficiency
\begin{equation}\label{Eq:pdfinal2}
\eta^{\rm inclined}(y) \approx \frac{1}{\sin \Theta}{\sqrt{\frac{2}{\pi}}\frac{\sigma_{\rm PD}}{ v_{\rm ion}}\frac{P}{h\nu w} \exp{\left(-\frac{2y^2}{w^2}\right)}}\,.
\end{equation}
We will not consider the extreme case of collinear beams ($\Theta=0$) here, as it is important for our purposes to be able to separate the neutral beam from the laser.   

In practice, one needs to know the overlapping densities of both beams at each position in the interaction region for precise calculations of the photodetachment efficiency. 
In the next sections we will present both numerical simulations based on the individual beam densities, as well as simplified calculations using the above formalism. We will  compare the results with experimental photodetachment efficiencies to extract the relevant parameters and highlight the potential of neutral beam creation using diode laser stacks.


\section{Experimental setup}\label{setup}
\subsection{Laser systems}

%
%
\begin{figure}[b]
\begin{center}
\includegraphics[width=8.6cm]{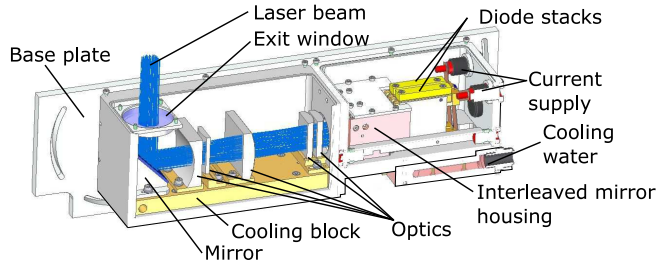} \caption{Schematic of the internal layout of the DILAS 808-nm-laser head, capable of delivering 2\,kW of continuous wave power.\label{Fig:LaserHead}}
\end{center}
\end{figure}
%

For our photodetachment studies we used two DILAS (Diode Laser Inc.) direct diode laser systems that were custom-built for this purpose. The lasers operate in fixed wavelength bands at $975\pm 5$\,nm and $808\pm 3$\,nm, where we have given each band center and full width at half maximum (FWHM). A schematic of one of the laser heads is shown in Fig.\,\ref{Fig:LaserHead}.  Each laser is composed of two water-cooled, vertically stacked diode bar arrays. 

The 975-nm-laser system was used to neutralize H$^-$. The laser is capable of delivering up to $1.4$~kW of continuous wave (CW) laser power from the two stacks, each of which consist of 12 diode bars. Each diode bar has a thickness (pitch) of 1.8\,mm and holds 19 individual emitters. Micro-optical lenses are used to collimate the highly divergent individual beamlets in both axes. Both stacks are optically combined by a set of refractive elements, creating a single combined beam (``interleaved stack'') with a pitch of 0.9\,mm. 
 The beam is then further shaped and focused by a series of standard optical elements. This arrangement results in a rectangular matrix of 19 $\times$ 24 individual beamlets at the laser output window. The optics are chosen such that the combined beamlets come to a focus $\sim$1.1\,m downstream of the laser exit. 

The newer 808-nm-laser system was used for C$^-$ photodetachment. The laser delivers up to $2.0$\,kW of CW power from two arrays with 15 diode bars each. The increased number of bars results in a slightly larger initial profile for the combined beam, compared to the 975-nm-laser, but other than that the interleaved optics and the beam geometry are very similar. 

\begin{figure}[t]
\begin{center}
\includegraphics[width=0.38\textwidth]{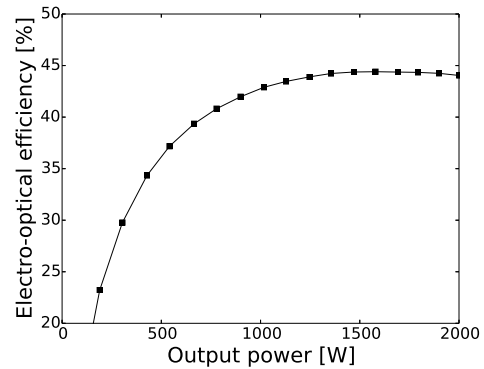} \caption{Measured electro-optical conversion efficiency of the 808-nm-diode laser system.\label{Fig:Efficiency}}
\end{center}
\end{figure}

Direct diode laser arrays are a very compact and reliable means to achieve high power densities. Electro-optical power conversion ratios of $60-65$\,\% can be reached under optimal conditions for state-of-the art material. Such high efficiencies make these laser systems particularly attractive for industrial applications where low power consumption is relevant. The 808-nm-system used in the present work achieved a conversion ratio of $45$\,\% under typical operating conditions (Fig.\,\ref{Fig:Efficiency}).
 
Furthermore, diode arrays often run maintenance-free for more than 10,000 hrs \cite{Dorsch1999}. In the present study the laser stacks were driven by power supplies manufactured by Amtron GmbH. Currents up to 100\,A and voltages of $\sim$40\,V were used. The Amtron power supplies allow for fast switching of the laser beam, with rise and fall times on the order of 20\,$\mu$s. Operating conditions ranging from CW to pulsed operation at frequencies up to 1\,kHz have been tested. 

%
%

\subsection{Laser beam characteristics}\label{laserbeam}
The diode laser stacks consist of an array of collimated laser diodes that can be modelled as an incoherent bundle of super-Gaussian beams (for specifics on diode laser characteristics and Gaussian beam propagation, see, e.g., \cite{Sun2015}). Specifically, while rays associated with the propagation direction of the individual beamlets obey the laws of geometrical optics, the beamlet profiles are obeying the laws of Gaussian optics (with the appropriate Gaussian beam quality factor $M^2$ \cite{Sun2015}). This explains how the beam appears box-like in the near field and Gaussian-like near and beyond the focus. The global waist is a compromise between the geometrical focus and the diffractive propagation of the individual beamlets.

%
\begin{figure}[t]
\begin{center}
\includegraphics[width=0.4\textwidth]{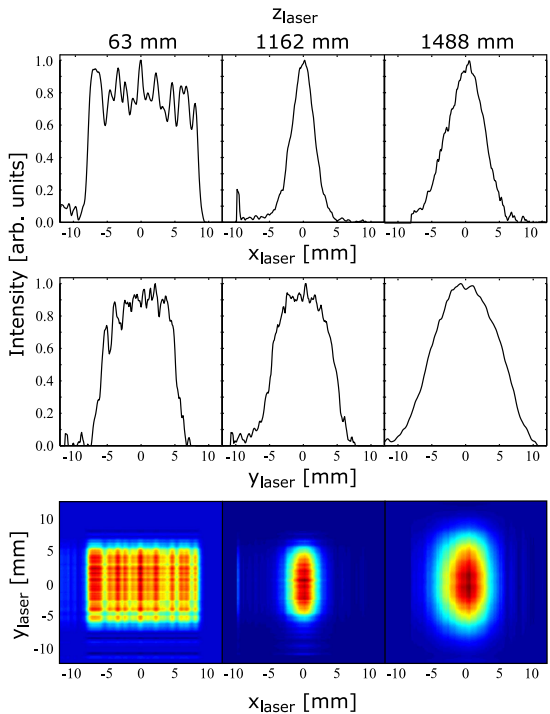} \caption{808-nm-laser profiles measured at three distances $z_\mathrm{\rm laser}$ from the laser exit window. The profiles were measured just above the laser diode threshold current of $\sim$10~A.\label{Fig:808Profile}}
\end{center}
\end{figure}

We measured the horizontal and vertical profile of the 808-nm-laser as a function of distance from the laser exit window, using the scanning knife edge technique \cite{Marshall2011}. Figure\,\ref{Fig:808Profile} shows the measured profiles at a selection of distances $z_{\rm laser}$ from the laser head. At $z_\mathrm{laser}$ = 63\,mm, the vertical profile lines are a diffuse image of the laser diode bars, which are focused by a series of lenses to an asymmetric spot of approximately 5\,mm $\times$ 10\,mm at $z_\mathrm{laser}$ = 1162\,mm. This focal length has been chosen to match the geometry of our photodetachment chamber (see Sec.\,\ref{Sec:pd-chamber}). As for most diode lasers, the divergence of the individual beamlets differs strongly in $x_{\rm laser}$- and $y_{\rm laser}$ direction (often referred to as the fast- and  slow axes, respectively). The individual microlenses cannot fully compensate for this effect and therefore the laser beam is not collimated and its beam quality is poor compared to typical single-emitter laser systems. However, since the most important parameter for our purposes is the power density and the overall vertical beam waist at the focus point, the limited beam quality is not a major concern. Of more practical impact is the development of side beams at high output power. These side lobes result from subtle changes in the emission profiles of the individual diodes that can occur when the cooling efficiency enters a critical regime. At high operating current, the divergence increases and overfills the slow axis collimating optics. In this case some of the emitted light can miss the microlens and end up outside the nominal beam. For our laser systems up to 15\,\% of the laser power in CW operation at maximum power ended up in side lobes and was dissipated in the vacuum chamber instead of leaving through the exit vacuum viewport.   

\begin{figure}[b]
\begin{center}
\includegraphics[width=0.4\textwidth]{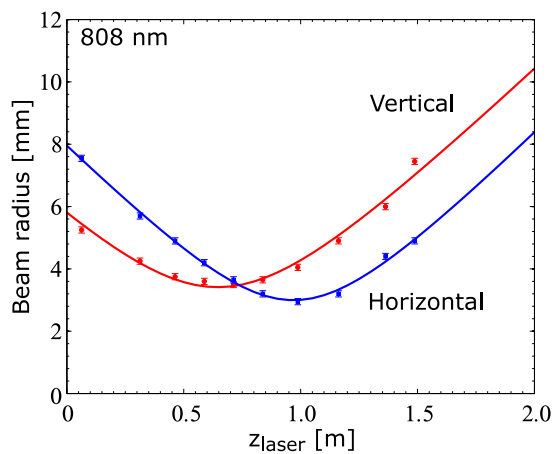} \caption{Measured beam radius (90\,\% transmission) of the 808-nm-laser system, as a function of distance from the laser exit window, in the vertical and horizontal planes.  Distributions are fitted with Eq.\,\ref{Eqn:BeamWaist}, the resulting fits are denoted by the solid curves. \label{Fig:CLaserRadius}}
\end{center}
\end{figure}

%
\begin{figure}[t]
\begin{center}
\includegraphics[width=0.4\textwidth]{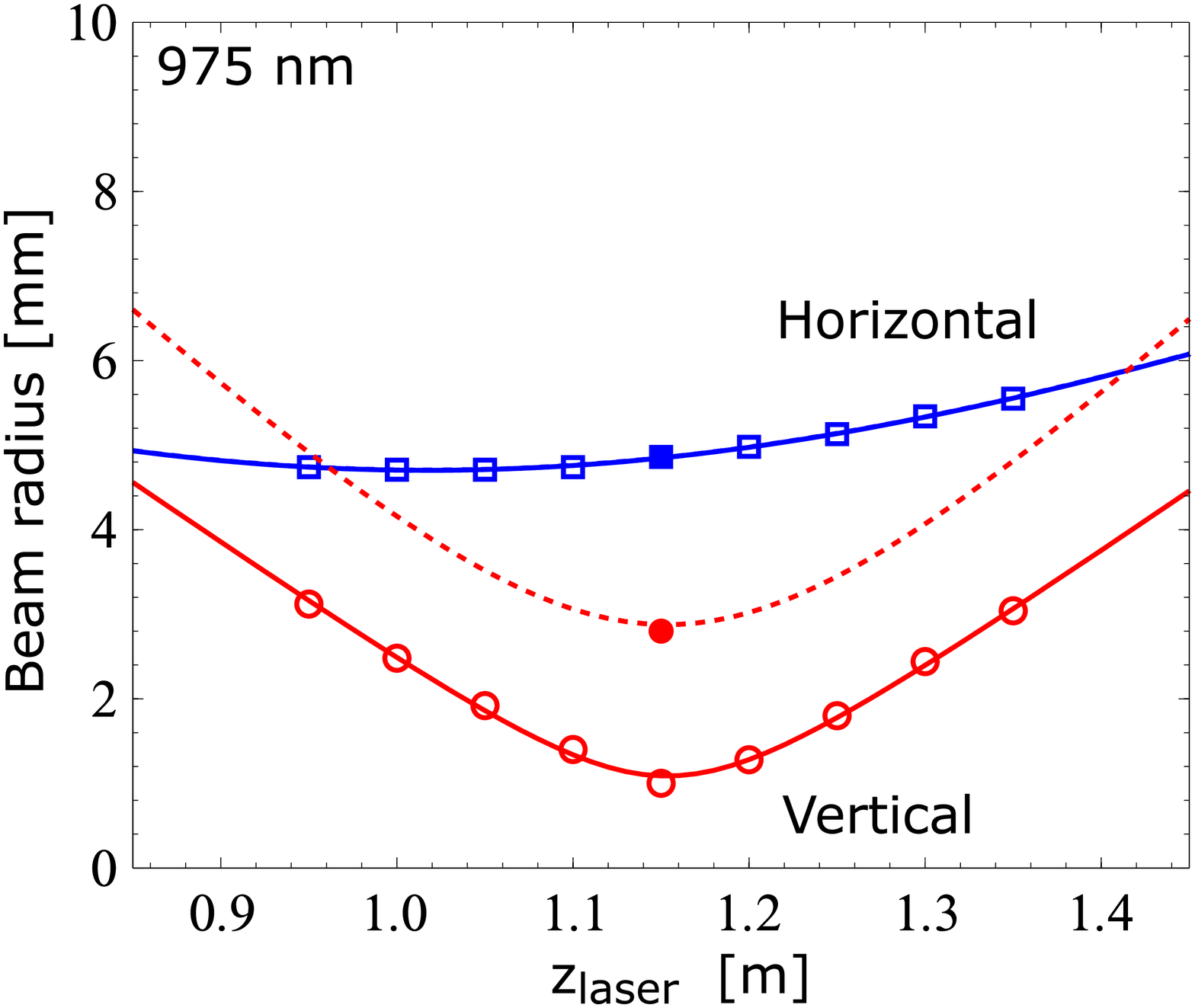} \caption{Laser beam radius (90\,\% transmission) for the 975-nm-laser as a function of distance from the laser head, in the vertical (circles) and horizontal plane (squares). Simulated profiles are denoted by hollow markers and the single measurement by the solid circle and solid square at $z_{\rm laser}=1150$\,mm. The simulated distributions were fitted using Eq.\ref{Eqn:BeamWaist}. The fit results are shown by the solid lines. The vertical profiles were then scaled to match the single measured profile in the focus (to account for beam broadening due to near-field non-linearities that are not included in the simulation), resulting in the dashed line. \label{Fig:LaserRadius}}
\end{center}
\end{figure}
%

\begin{figure*}[t]
\begin{center}
\includegraphics[width=0.75\textwidth]{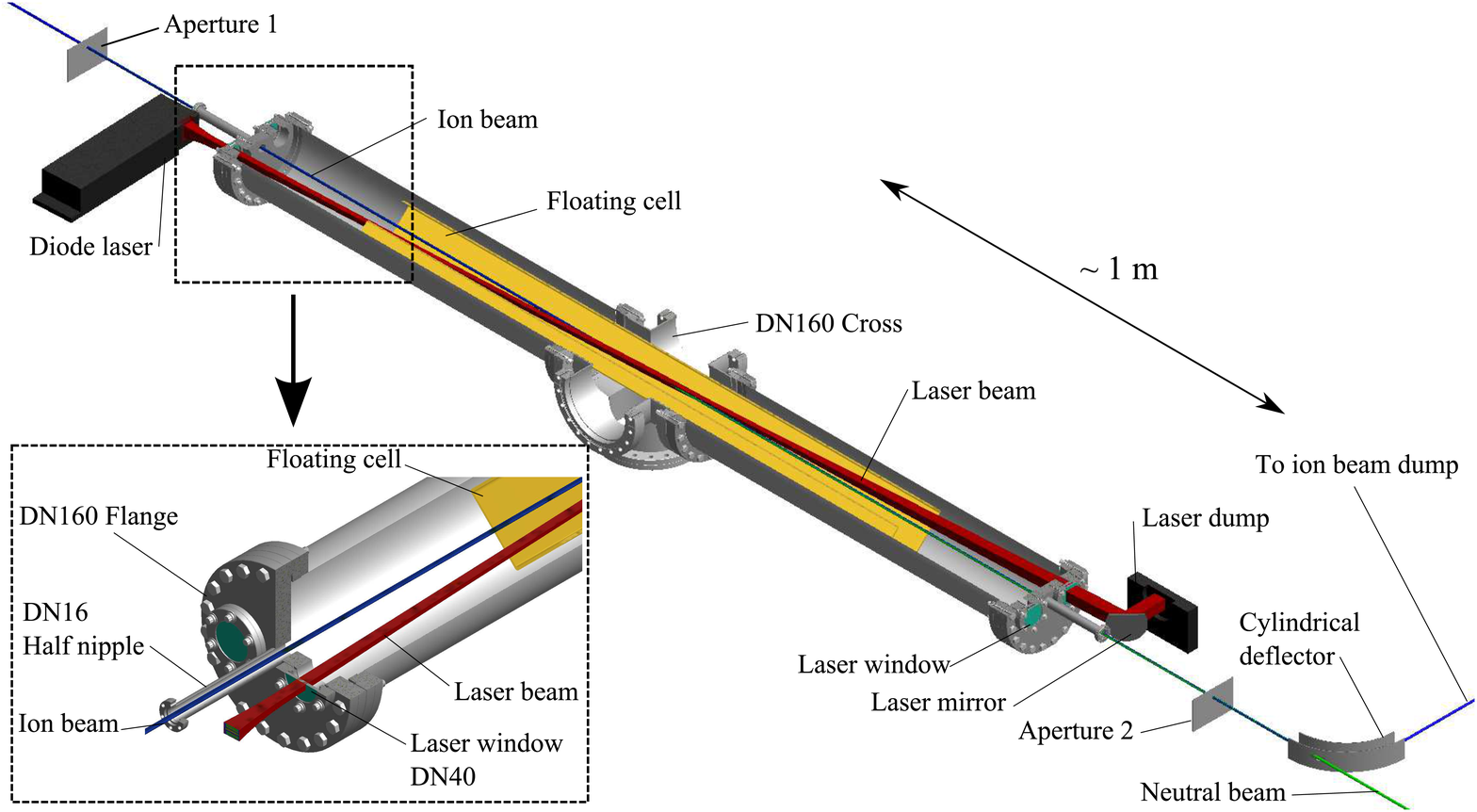} \caption{Schematic overview of the photodetachment region.\label{Fig:PD_Chamber}}
\end{center}
\end{figure*}

For simulations of the overall photodetachment efficiency, we needed a quantifiable measure of the dimension of the combined laser beam at different positions along the propagation axis. To this end we fitted the measured horizontal and vertical laser profiles as an imperfect Gaussian beam according to \cite{Sun2015}
\begin{equation}\label{Eqn:BeamWaist}
w\left(z_\mathrm{laser}\right) = w_0 \sqrt{1 +  \left[\frac{M^2\,\lambda \left(z_\mathrm{laser}-z_0\right) }{\pi w_0^2}\right]^2}
\end{equation}
where $w(z_{\rm laser})$ is the beam waist at position $z_\mathrm{laser}$, for a beam with a Gaussian beam quality factor $M^2$, wavelength $\lambda$, focus location $z_0$ with respect to the laser exit window, and minimum beam waist $w_0$. While this approximation is clearly not very good at the laser exit, it improves significantly close to the more interesting region of the global beam focus. We assumed independent diffractive propagation in the vertical and horizontal directions, which we denote $x_{\rm laser}$ and $y_{\rm laser}$, respectively, in the reference frame of the laser. In Fig.\,\ref{Fig:CLaserRadius} the measured beam radius along the propagation axis is plotted for the 808-nm-laser, together with a fit using Eq.\,\ref{Eqn:BeamWaist}. The fair agreement shows that the overall propagation of the combined beamlets can be approximated by a single Gaussian beam with a large quality correction factor of $M^2 \sim 100$.

\begin{table}[b!]
\begin{center}
\caption{Effective Gaussian beam parameters for both laser systems. The parameters have been derived using Eq.\,\ref{Eqn:BeamWaist}.}
\begin{tabular}{|l|l|c|c|c|}
\hline
\multicolumn{2}{|c|}{ }     &\quad  $w_0$ [mm] \quad &\quad $M^2$\qquad& \quad $z_0$ [m]\quad \\
\hline
\hline
808\,nm-laser \hspace*{0.3cm} & horizontal & $3.0$                   & 88 & $0.97$ \\
\cline{2-5}
               &  vertical  & $3.4$                   & 97 & $0.65$\\ 
\hline
\hline
975\,nm-laser  \qquad & horizontal & $4.7$                   & 135 & $1.02$\\
\cline{2-5}
               & vertical   & $2.9$                   & $182$ & $1.15$\\ 
\hline
\end{tabular}\label{Tab:gaussians}
\end{center}
\end{table}

We did not carry out detailed profile measurements for the 975-nm-laser system. Here we had to rely on optical simulations performed using the ZEMAX optical design suite \cite{zemax}. The results of the simulation in the vicinity of the laser focus at $\sim$1.1\,m are shown in Fig.\,\ref{Fig:LaserRadius}. The simulations did not take into account near field non-linearities that deform the laser beam along individual laser bars (resulting in the so-called ``smile'' \cite{wang09}). For our laser system these aberrations will result in a linear broadening of the vertical profiles for all distances. To account for the broadening caused by this effect, we compared the simulated profiles to a single measurement that was carried out at the laser beam focus. Then we scaled the simulated vertical profiles up to match the measurement. The results of the fits,  given as effective Gaussian beam parameters for both laser systems, are listed in Tab.\,\ref{Tab:gaussians}.

\subsection{Photodetachment chamber}\label{Sec:pd-chamber}
The photodetachment chamber depicted in Fig.\,\ref{Fig:PD_Chamber} was initially designed to facilitate efficient neutral beam creation for an ion-neutral merged-beams experiment \cite{science2010, bruhns2010, ocon15}. At the present time, several versions of this chamber are in operation at Columbia University and the Max-Planck-Institut f\"ur Kernphysik (MPIK) in Heidelberg, Germany. Except for minor adaptations, the general features and dimensions remain unchanged.

The photodetachment takes place in an $\sim$2\,m long vacuum chamber that consists of a central Conflat (CF) DN160 six-way cross and two straight sections with a length of $\sim$80\,cm each. Custom-built end flanges (see inset in Fig.\,\ref{Fig:PD_Chamber}) allow for the injection of the negative ion beam through small DN16 half nipples. The laser enters and exits the vacuum through DN40 windows and intersects the ion beam at a small angle of $\Theta=2.7\,^\circ$ in the central cross. The small intersection angle extends the overlap region and thereby boosts the photodetachment efficiency -- which scales with $1/\sin{\Theta}$ -- by a factor of $\sim$21 (compared to a crossed-beam arrangement at right angles). Due to differences in the laser head geometry, only the 975-nm-laser head could be coupled directly into the photodetachment chamber, whereas the 808-nm-system required the inclusion of an angled mirror between the laser head and the vacuum entry window. This influenced the laser path length from the head to the central chamber. Path lengths from the exit window of the laser head to the middle of the photodetachment chamber of $\sim$1.05\,m and $\sim$1.20\,m were measured, for the 975- and 808-nm-lasers, respectively.  

To collimate the ion and neutral beams, the photodetachment region is bounded by two circular apertures of 5\,mm diameter that are separated by 280\,cm (for the H$^-$ measurements) or 307\,cm (for the C$^-$ measurements). 

Behind the second aperture the remaining ion beam can be separated from the neutrals using electric or magnetic fields. In our experimental setups, we have used both an electrostatic quadrupole and a simple cylindrical deflector (similar to that described by Kreckel et al. \cite{Kreckel2010}, but with a hole in the rear plate). 
The current of the deflected ion beam is measured in a Faraday cup. The neutral particle flux is measured via calibrated secondary electron emission in a modified Faraday cup that features separate electrodes for neutral impact and for collection of the secondary electrons.

In order to fine-tune the neutral beam energy, we installed a floating cell with a length of $\sim$1.4\,m in the photodetachment chamber. By applying a negative (positive) voltage to this electrode, the ions are decelerated (accelerated) at the entrance and accelerated (decelerated) back to their original kinetic energy at the exit of the cell. The neutrals that are created in the center of the floating cell, however, will remain at the lower (higher) kinetic energy and continue to fly ballistically. Hence with the use of the floating cell we can adjust the neutral beam energy without having to change the acceleration potential of the ion source or the ion beam optics.

\subsection{Ion beam characteristics}\label{ionbeam}
For the present study, we have used H$^-$ and C$^-$beams at 10\,keV and 28\,keV, respectively. While the H$^-$ ion beam was produced in an off-axis duoplasmatron source, the C$^-$ ions were extracted from a conventional sputter ion source. The ions were mass-selected by a Wien filter and then focused and guided into the photodetachment region by electrostatic ion optical elements. A right angle deflection was introduced between each ion source and the corresponding photodetachment chamber to prevent gas and neutralized atoms from the source region from making their way into the experimental section.

In the straight photodetachment chamber two circular apertures are used to limit the beam emittance. For the H$^-$ case we can interpret the divergence angle and beam diameter permitted by the two 5\,mm apertures at 280\,cm distance as a maximum transmitted area in phase space of $A= \pi \epsilon=8.9$\,mm\,mrad (where $\epsilon$ stands for the beam emittance). For the C$^-$ beam, the apertures were placed 307\,cm  apart, resulting in a slightly smaller phase space area of $A=8.1$\,mm\,mrad. One should keep in mind, though, that these limits apply to transmitted beams on an unperturbed ballistic trajectory. In reality, the ions are propagating under the influence of mutual space charge repulsion, while the neutrals are not subject to space charge effects. This leads to differences between the fraction of ions that were neutralized by the laser and the ratio of transmitted neutrals to transmitted ions. We will quantify these effects in the next section. 

We typically started out with ion beam currents measured in front of the photodetachment section that were on the order of $\sim$5\,$\mu$A for H$^-$ and $\sim$1\,$\mu$A for C$^-$. We measured the transmitted ion beam current after each aperture. These parameters together with SIMION \cite{simionweb} simulations and the assumption of circular initial beams allowed us to reconstruct the ion beam waist in the middle of the photodetachment chamber. We compared the simulations with beam profile measurements using wire scanners that were positioned downstream of the photodetachment region and achieved very good agreement. For the H$^-$ beam, we derived a Gaussian beam radius of 1.3\,mm (1/e$^2$) in the center of the photodetachment chamber; and for C$^-$ we derived 1.1\,mm. These values were used in combination with the laser beam profiles to simulate the overall photodetachment efficiencies.

\begin{figure}[b]
\begin{center}
\includegraphics[width=0.4\textwidth]{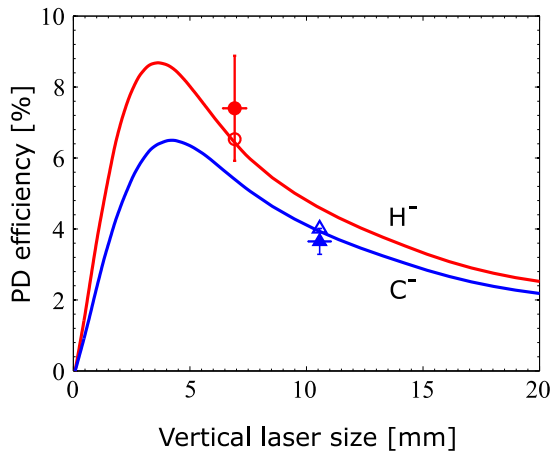}
\caption{Photodetachment (PD) efficiencies for 10~keV H$^-$ (red circles) and 28~keV C$^-$ ions (blue triangles). The solid symbols show the experimental results. The open symbols are the result of Monte Carlo simulations taking into account the ion and neutral beam profiles in the overlap region. The solid lines depict simplified calculations assuming ion beams with a fixed Gaussian profile and homogeneous cylindrical laser beams. The power of the  975-nm- and 808-nm-laser in normal operation was 1.4\,kW and 1.8\,kW, respectively. The vertical laser beam size is given by the distance in $y_{\rm laser}$ that contains 90\% of the power for the Monte Carlo simulation and the experimental beam profiles, while for the simplified simulation it represents the beam diameter that contains 90\% of the laser power. \label{Fig:GaussLaser}}
\end{center}
\end{figure}

\section{Photodetachment efficiencies}
The total photodetachment efficiency was calculated from Monte Carlo simulations based on the ion and laser beam profiles derived above (Sec.\,\ref{laserbeam} and \ref{ionbeam}) and compared to direct  measurements of the ion and neutral signals. Our simulations incorporated the change in the horizontal and vertical distributions of both laser and ion beams within the overlap region. Ions were flown at their nominal kinetic energy, starting from a plane upstream of the first aperture. The individual transversal positions and velocity angles were distributed according to a Gaussian ion beam distribution with a 10\,eV FWHM energy spread, which is common for the ion sources used here. Typical simulations consisted of flying 10,000 ions over a photon-field divided into cells with a volume of $640\times30\times30$ $\mu$m$^3$ each. The photon flux in each cell was derived from the measured laser profiles. For each individual trajectory the averaged photon flux and flight time through the laser field were recorded, from which the photodetachment probability for each ion was calculated. We assumed cross sections of $3.8\times 10^{-17}$\,cm$^2$ \cite{gene15} for photodetachment of H$^-$ at 975\,nm and $1.5\times 10^{-17}$\,cm$^2$ \cite{seman1962} for C$^-$ at 808\,nm. The simulated photodetachment efficiencies for our experimental setups are shown in Fig.\,\ref{Fig:GaussLaser} as hollow symbols. For the H$^-$ and C$^-$ beams, the Monte Carlo simulation yielded photodetachment efficiencies of 6.5\,\% and 4.0\,\%, respectively. Also shown in the plot are the experimentally determined efficiencies of $\sim$7.4\,\% for H$^-$ and $\sim$3.7\,\% for C$^-$ (solid symbols). The experimental values are based on measurements of the ion current reduction in the Faraday cup that collects the deflected ions downstream of the neutralization chamber. 

To come to a more systematic understanding of the efficiencies for different matchings of the ion and laser beam overlap, we carried out simplified simulations that are depicted by the solid lines. In this case we used constant Gaussian ion beam waists of 1.3\,mm for H$^-$ and 1.1\,mm for C$^-$ in the center of the overlap region, as determined for the optimal geometric transmittance of the beam through the photodetachment chamber. The laser beam was represented by a simple circular profile with a homogeneous photon density. For Fig.\,\ref{Fig:GaussLaser} we consistently use the 90\,\% transmitted intensity boundary as the vertical laser beam diameter. The comparison of this simplified calculation to the more sophisticated Monte Carlo approach shows very good agreement. The simple simulation allows us to estimate the ideal vertical laser beam size for our given ion beam profiles (remember that the horizontal laser profile basically does not matter, as the laser and ion beams intersect in the horizontal plane). The simulated curve peaks at a value of 9.2\,\% for H$^-$ and 6.2\,\% for C$^-$. This shows that we could gain an additional factor of $\sim$1.25 for H$^-$ if we could bring the laser to a smaller focus in the overlap region, while for C$^-$ we could gain a factor of $\sim$1.68. The reason for the sub-optimal configuration in the latter case is due to the fact that the laser axes of the new 808-nm-system are rotated by 90$^\circ$ compared to the 975-nm-system, resulting in a beam profile that is larger in vertical than in horizontal direction. For optimal neutral gain the axes of the 808-nm-system should be interchanged. In the neutral beam setup that is currently under construction at MPIK, the laser and ion beams will be intersecting in the vertical plane, to make full use of the laser power. However, for the sake of simplicity, the present measurements were carried out in the same horizontal chamber geometry as the 975-nm-laser, only the distances of the circular apertures were modified. Consequently, we expect to achieve approximately a factor of $\sim$1.68 gain in efficiency for C$^-$ at the neutral beam setup at MPIK.

For our merged beams experiments \cite{science2010,bruhns2010a,miller2011,miller2012}, we simultaneously measured the neutral flux exiting the photodetachment region through the second aperture and the reduction in the transmitted ion beam current, caused by the laser. For the neutral flux determination we used a calibrated cup in which we recorded the secondary electron emission off a metal surface. We noticed a clear difference in the ratio of the transmitted neutral and ion currents compared to the fraction of ions that were photodetached by the laser. The reason for this difference is that the ion beam is subjected to space charge while the neutrals are not. Therefore, the ion and neutral beams have quite different emittances, with the neutral beam effectively being more collimated. That also means that a part of the ion beam that will not make it through the second aperture still contributes to neutral beam creation. We simulated this effect with extensive SIMION calculations \cite{simionweb} including beam repulsion and using the known ion beam geometry. From the difference in transmission with and without beam repulsion we calculated an enhancement factor of 1.29 for the transmitted neutral-to-ion ratio compared to the photodetachment efficiency for H$^-$. This agreed very well with the measured neutral-to-ion ratio of $\sim$9.5\%. For the heavier C$^-$ system space charge effects are less pronounced and the calculated enhancement factor is 1.14. Note that this does not mean that the photodetachment is more efficient. In fact, the increased transmitted neutral-to-ion ratio is an experimental artifact of our setup, where the photodetachment occurs between the collimating apertures.

\section{Conclusion and Perspectives}\label{conclusion}
We have shown that photodetachment using direct diode laser stacks is an efficient technique for the creation of fast beams of neutral atoms. Neutralization efficiencies of $\sim$7.4\,\% and $\sim$3.7\,\% were achieved in an inclined beams configuration for H$^-$ and C$^-$ beams at kinetic energies of 10\,keV and 28\,keV, respectively. By using a small intersection angle between the ion and laser beams, we were able to increase the overlap length and boost the photodetachment efficiency by a factor of $\sim$21 compared to a crossed beams arrangement at right angles. The diode laser setups that were used here provide very stable power levels and require little maintenance, which is a major advantage for neutral beam creation at complex experiments like the future neutral beamline of the Cryogenic Storage Ring (CSR) \cite{vonhahn11} at MPIK in Heidelberg. Our simulations show that an ideal matching of the ion and laser beam profiles would allow us to reach $\sim$9.2\,\% photodetachment efficiency for H$^-$ at 10\,keV and $\sim$6.2\,\% for C$^-$ at 28\,keV. It should be noted that at lower kinetic energies even higher neutralization fractions could be achieved, as the photodetachment efficiency scales with $1/\sqrt{E_{\rm ion}}$. The focus of the present study has been the creation of fast well-defined neutral ground-term atomic beams for merged beams studies of astrophysical processes. However, the same technique could be applied to neutralize the majority of stable atomic anions in the periodic table.


\begin{acknowledgments}
Work at Columbia University was supported, in part, by the NSF Divisions of Chemistry and of Astronomical Sciences. A.O., F.G., and H.K. were supported by the European Research Council under Grant Agreement No.\,StG 307163 and by the Max Planck Society. H.B. was supported in part by the German academic exchange service DAAD. X.U. acknowledges support from the Fund for Scientific Research - FNRS and the IISN under Contract No.\,4.4504.10.
\end{acknowledgments}

\bibliographystyle{apsrev}
\bibliography{PD_Bib}

\begin{thebibliography}{39}
\expandafter\ifx\csname natexlab\endcsname\relax\def\natexlab#1{#1}\fi
\expandafter\ifx\csname bibnamefont\endcsname\relax
  \def\bibnamefont#1{#1}\fi
\expandafter\ifx\csname bibfnamefont\endcsname\relax
  \def\bibfnamefont#1{#1}\fi
\expandafter\ifx\csname citenamefont\endcsname\relax
  \def\citenamefont#1{#1}\fi
\expandafter\ifx\csname url\endcsname\relax
  \def\url#1{\texttt{#1}}\fi
\expandafter\ifx\csname urlprefix\endcsname\relax\def\urlprefix{URL }\fi
\providecommand{\bibinfo}[2]{#2}
\providecommand{\eprint}[2][]{\url{#2}}

\bibitem[{\citenamefont{{Wooden} et~al.}(2004)\citenamefont{{Wooden},
  {Charnley}, and {Ehrenfreund}}}]{wooden04}
\bibinfo{author}{\bibfnamefont{D.~H.} \bibnamefont{{Wooden}}},
  \bibinfo{author}{\bibfnamefont{S.~B.} \bibnamefont{{Charnley}}},
  \bibnamefont{and}
  \bibinfo{author}{\bibfnamefont{P.}~\bibnamefont{{Ehrenfreund}}},
  \emph{\bibinfo{title}{{Composition and evolution of interstellar clouds, {\rm
  in Comets II}}}} (\bibinfo{publisher}{University of Arizona Press, Tucson},
  \bibinfo{year}{2004}), pp. \bibinfo{pages}{33--66}.

\bibitem[{\citenamefont{{Wakelam} et~al.}(2009)\citenamefont{{Wakelam},
  {Loison}, {Herbst}, {Talbi}, {Quan}, and {Caralp}}}]{wakelam09}
\bibinfo{author}{\bibfnamefont{V.}~\bibnamefont{{Wakelam}}},
  \bibinfo{author}{\bibfnamefont{J.-C.} \bibnamefont{{Loison}}},
  \bibinfo{author}{\bibfnamefont{E.}~\bibnamefont{{Herbst}}},
  \bibinfo{author}{\bibfnamefont{D.}~\bibnamefont{{Talbi}}},
  \bibinfo{author}{\bibfnamefont{D.}~\bibnamefont{{Quan}}}, \bibnamefont{and}
  \bibinfo{author}{\bibfnamefont{F.}~\bibnamefont{{Caralp}}},
  \bibinfo{journal}{Astron. Astrophys.} \textbf{\bibinfo{volume}{495}},
  \bibinfo{pages}{513} (\bibinfo{year}{2009}).

\bibitem[{\citenamefont{{Snow} and {Bierbaum}}(2008)}]{snow08}
\bibinfo{author}{\bibfnamefont{T.~P.} \bibnamefont{{Snow}}} \bibnamefont{and}
  \bibinfo{author}{\bibfnamefont{V.~M.} \bibnamefont{{Bierbaum}}},
  \bibinfo{journal}{Annu. Rev. Anal. Chem.} \textbf{\bibinfo{volume}{1}},
  \bibinfo{pages}{229} (\bibinfo{year}{2008}).

\bibitem[{\citenamefont{{Urbain} et~al.}(1991)\citenamefont{{Urbain}, {Cornet},
  {Brouillard}, and {Giusti-Suzor}}}]{urbain91}
\bibinfo{author}{\bibfnamefont{X.}~\bibnamefont{{Urbain}}},
  \bibinfo{author}{\bibfnamefont{A.}~\bibnamefont{{Cornet}}},
  \bibinfo{author}{\bibfnamefont{F.}~\bibnamefont{{Brouillard}}},
  \bibnamefont{and}
  \bibinfo{author}{\bibfnamefont{A.}~\bibnamefont{{Giusti-Suzor}}},
  \bibinfo{journal}{Phys. Rev. Lett.} \textbf{\bibinfo{volume}{66}},
  \bibinfo{pages}{1685} (\bibinfo{year}{1991}).

\bibitem[{\citenamefont{{Daranlot} et~al.}(2011)\citenamefont{{Daranlot},
  {Jorfi}, {Xie}, {Bergeat}, {Costes}, {Caubet}, {Xie}, {Guo}, {Honvault}, and
  {Hickson}}}]{daranlot11}
\bibinfo{author}{\bibfnamefont{J.}~\bibnamefont{{Daranlot}}},
  \bibinfo{author}{\bibfnamefont{M.}~\bibnamefont{{Jorfi}}},
  \bibinfo{author}{\bibfnamefont{C.}~\bibnamefont{{Xie}}},
  \bibinfo{author}{\bibfnamefont{A.}~\bibnamefont{{Bergeat}}},
  \bibinfo{author}{\bibfnamefont{M.}~\bibnamefont{{Costes}}},
  \bibinfo{author}{\bibfnamefont{P.}~\bibnamefont{{Caubet}}},
  \bibinfo{author}{\bibfnamefont{D.}~\bibnamefont{{Xie}}},
  \bibinfo{author}{\bibfnamefont{H.}~\bibnamefont{{Guo}}},
  \bibinfo{author}{\bibfnamefont{P.}~\bibnamefont{{Honvault}}},
  \bibnamefont{and} \bibinfo{author}{\bibfnamefont{K.~M.}
  \bibnamefont{{Hickson}}}, \bibinfo{journal}{Science}
  \textbf{\bibinfo{volume}{334}}, \bibinfo{pages}{1538} (\bibinfo{year}{2011}).

\bibitem[{\citenamefont{O'Connor et~al.}(2015)\citenamefont{O'Connor, Urbain,
  St\"utzel, Miller, de~Ruette, Garrido, and Savin}}]{ocon15}
\bibinfo{author}{\bibfnamefont{A.~P.} \bibnamefont{O'Connor}},
  \bibinfo{author}{\bibfnamefont{X.}~\bibnamefont{Urbain}},
  \bibinfo{author}{\bibfnamefont{J.}~\bibnamefont{St\"utzel}},
  \bibinfo{author}{\bibfnamefont{K.~A.} \bibnamefont{Miller}},
  \bibinfo{author}{\bibfnamefont{N.}~\bibnamefont{de~Ruette}},
  \bibinfo{author}{\bibfnamefont{M.}~\bibnamefont{Garrido}}, \bibnamefont{and}
  \bibinfo{author}{\bibfnamefont{D.~W.} \bibnamefont{Savin}},
  \bibinfo{journal}{ApJS} \textbf{\bibinfo{volume}{219}}
  (\bibinfo{year}{2015}).

\bibitem[{\citenamefont{{Hirsch} et~al.}(1997)\citenamefont{{Hirsch}, {Yu},
  {Buczkowski}, {Myers}, and {Richards-Babb}}}]{hirsch97}
\bibinfo{author}{\bibfnamefont{L.~S.} \bibnamefont{{Hirsch}}},
  \bibinfo{author}{\bibfnamefont{Z.}~\bibnamefont{{Yu}}},
  \bibinfo{author}{\bibfnamefont{S.~L.} \bibnamefont{{Buczkowski}}},
  \bibinfo{author}{\bibfnamefont{T.~H.} \bibnamefont{{Myers}}},
  \bibnamefont{and} \bibinfo{author}{\bibfnamefont{M.~R.}
  \bibnamefont{{Richards-Babb}}}, \bibinfo{journal}{J. Electron. Mater.}
  \textbf{\bibinfo{volume}{26}}, \bibinfo{pages}{534} (\bibinfo{year}{1997}).

\bibitem[{\citenamefont{{Hemsworth} et~al.}(2009)\citenamefont{{Hemsworth},
  {Decamps}, {Graceffa}, {Schunke}, {Tanaka}, {Dremel}, {Tanga}, {DeEsch},
  {Geli}, {Milnes} et~al.}}]{hemsworth09}
\bibinfo{author}{\bibfnamefont{R.}~\bibnamefont{{Hemsworth}}},
  \bibinfo{author}{\bibfnamefont{H.}~\bibnamefont{{Decamps}}},
  \bibinfo{author}{\bibfnamefont{J.}~\bibnamefont{{Graceffa}}},
  \bibinfo{author}{\bibfnamefont{B.}~\bibnamefont{{Schunke}}},
  \bibinfo{author}{\bibfnamefont{M.}~\bibnamefont{{Tanaka}}},
  \bibinfo{author}{\bibfnamefont{M.}~\bibnamefont{{Dremel}}},
  \bibinfo{author}{\bibfnamefont{A.}~\bibnamefont{{Tanga}}},
  \bibinfo{author}{\bibfnamefont{H.~P.~L.} \bibnamefont{{DeEsch}}},
  \bibinfo{author}{\bibfnamefont{F.}~\bibnamefont{{Geli}}},
  \bibinfo{author}{\bibfnamefont{J.}~\bibnamefont{{Milnes}}},
  \bibnamefont{et~al.}, \bibinfo{journal}{Nucl. Fusion}
  \textbf{\bibinfo{volume}{49}}, \bibinfo{eid}{045006} (\bibinfo{year}{2009}).

\bibitem[{\citenamefont{{Kavari} and {Crowley}}(2010)}]{kovari10}
\bibinfo{author}{\bibfnamefont{M.}~\bibnamefont{{Kavari}}} \bibnamefont{and}
  \bibinfo{author}{\bibfnamefont{B.}~\bibnamefont{{Crowley}}},
  \bibinfo{journal}{Fusion Eng. Des.} \textbf{\bibinfo{volume}{85}},
  \bibinfo{pages}{745} (\bibinfo{year}{2010}).

\bibitem[{\citenamefont{{Fink} and {Frank}}(1975)}]{fink75}
\bibinfo{author}{\bibfnamefont{J.}~\bibnamefont{{Fink}}} \bibnamefont{and}
  \bibinfo{author}{\bibfnamefont{A.}~\bibnamefont{{Frank}}},
  \emph{\bibinfo{title}{Photodetachment of electrons from negative ions in a
  200 keV deuterium beam source, {\rm Lawrence Livermore Natl. Lab.,
  UCRL-16844}}} (\bibinfo{year}{1975}).

\bibitem[{\citenamefont{{van Zyl} et~al.}(1976)\citenamefont{{van Zyl},
  {Utterback}, and {Amme}}}]{vanzyl76}
\bibinfo{author}{\bibfnamefont{B.}~\bibnamefont{{van Zyl}}},
  \bibinfo{author}{\bibfnamefont{N.~G.} \bibnamefont{{Utterback}}},
  \bibnamefont{and} \bibinfo{author}{\bibfnamefont{R.~C.}
  \bibnamefont{{Amme}}}, \bibinfo{journal}{Rev. Sci. Instrum.}
  \textbf{\bibinfo{volume}{47}}, \bibinfo{pages}{814} (\bibinfo{year}{1976}).

\bibitem[{\citenamefont{{Havener} et~al.}(1989)\citenamefont{{Havener}, {Huq},
  {Krause}, {Schulz}, and {Phaneuf}}}]{havener89}
\bibinfo{author}{\bibfnamefont{C.~C.} \bibnamefont{{Havener}}},
  \bibinfo{author}{\bibfnamefont{M.~S.} \bibnamefont{{Huq}}},
  \bibinfo{author}{\bibfnamefont{H.~F.} \bibnamefont{{Krause}}},
  \bibinfo{author}{\bibfnamefont{P.~A.} \bibnamefont{{Schulz}}},
  \bibnamefont{and} \bibinfo{author}{\bibfnamefont{R.~A.}
  \bibnamefont{{Phaneuf}}}, \bibinfo{journal}{Phys. Rev. A}
  \textbf{\bibinfo{volume}{39}}, \bibinfo{pages}{1725} (\bibinfo{year}{1989}).

\bibitem[{\citenamefont{{Bruhns} et~al.}(2008)\citenamefont{{Bruhns},
  {Kreckel}, {Savin}, {Seely}, and {Havener}}}]{bruhns08}
\bibinfo{author}{\bibfnamefont{H.}~\bibnamefont{{Bruhns}}},
  \bibinfo{author}{\bibfnamefont{H.}~\bibnamefont{{Kreckel}}},
  \bibinfo{author}{\bibfnamefont{D.~W.} \bibnamefont{{Savin}}},
  \bibinfo{author}{\bibfnamefont{D.~G.} \bibnamefont{{Seely}}},
  \bibnamefont{and} \bibinfo{author}{\bibfnamefont{C.~C.}
  \bibnamefont{{Havener}}}, \bibinfo{journal}{Phys. Rev. A}
  \textbf{\bibinfo{volume}{77}}, \bibinfo{eid}{064702} (\bibinfo{year}{2008}).

\bibitem[{\citenamefont{{Dragani{\'c}}
  et~al.}(2011)\citenamefont{{Dragani{\'c}}, {Seely}, and
  {Havener}}}]{draganic11}
\bibinfo{author}{\bibfnamefont{I.~N.} \bibnamefont{{Dragani{\'c}}}},
  \bibinfo{author}{\bibfnamefont{D.~G.} \bibnamefont{{Seely}}},
  \bibnamefont{and} \bibinfo{author}{\bibfnamefont{C.~C.}
  \bibnamefont{{Havener}}}, \bibinfo{journal}{Phys. Rev. A}
  \textbf{\bibinfo{volume}{83}}, \bibinfo{eid}{054701} (\bibinfo{year}{2011}).

\bibitem[{\citenamefont{Andersen et~al.}(1999)\citenamefont{Andersen, Haugen,
  and Hotop}}]{andersen99}
\bibinfo{author}{\bibfnamefont{T.}~\bibnamefont{Andersen}},
  \bibinfo{author}{\bibfnamefont{H.~K.} \bibnamefont{Haugen}},
  \bibnamefont{and} \bibinfo{author}{\bibfnamefont{H.}~\bibnamefont{Hotop}},
  \bibinfo{journal}{J. Phys. Chem. Ref. Data} \textbf{\bibinfo{volume}{28}}
  (\bibinfo{year}{1999}).

\bibitem[{\citenamefont{{Andersen}}(2004)}]{andersen04}
\bibinfo{author}{\bibfnamefont{T.}~\bibnamefont{{Andersen}}},
  \bibinfo{journal}{Phys. Rep.} \textbf{\bibinfo{volume}{394}},
  \bibinfo{pages}{157} (\bibinfo{year}{2004}).

\bibitem[{\citenamefont{{Wigner}}(1948)}]{wigner1948}
\bibinfo{author}{\bibfnamefont{E.~P.} \bibnamefont{{Wigner}}},
  \bibinfo{journal}{Phys. Rev.} \textbf{\bibinfo{volume}{73}},
  \bibinfo{pages}{1002} (\bibinfo{year}{1948}).

\bibitem[{\citenamefont{{Scheer} et~al.}(1998)\citenamefont{{Scheer},
  {Bilodeau}, {Brodie}, and {Haugen}}}]{scheer98}
\bibinfo{author}{\bibfnamefont{M.}~\bibnamefont{{Scheer}}},
  \bibinfo{author}{\bibfnamefont{R.~C.} \bibnamefont{{Bilodeau}}},
  \bibinfo{author}{\bibfnamefont{C.~A.} \bibnamefont{{Brodie}}},
  \bibnamefont{and} \bibinfo{author}{\bibfnamefont{H.~K.}
  \bibnamefont{{Haugen}}}, \bibinfo{journal}{\pra}
  \textbf{\bibinfo{volume}{58}}, \bibinfo{pages}{2844} (\bibinfo{year}{1998}).

\bibitem[{\citenamefont{{Takao} et~al.}(2007)\citenamefont{{Takao}, {Jinno},
  {Hanada}, {Goto}, {Oshikiri}, {Okuno}, {Tanuma}, {Azuma}, and
  {Shiromaru}}}]{takao07}
\bibinfo{author}{\bibfnamefont{T.}~\bibnamefont{{Takao}}},
  \bibinfo{author}{\bibfnamefont{S.}~\bibnamefont{{Jinno}}},
  \bibinfo{author}{\bibfnamefont{K.}~\bibnamefont{{Hanada}}},
  \bibinfo{author}{\bibfnamefont{M.}~\bibnamefont{{Goto}}},
  \bibinfo{author}{\bibfnamefont{K.}~\bibnamefont{{Oshikiri}}},
  \bibinfo{author}{\bibfnamefont{K.}~\bibnamefont{{Okuno}}},
  \bibinfo{author}{\bibfnamefont{H.}~\bibnamefont{{Tanuma}}},
  \bibinfo{author}{\bibfnamefont{T.}~\bibnamefont{{Azuma}}}, \bibnamefont{and}
  \bibinfo{author}{\bibfnamefont{H.}~\bibnamefont{{Shiromaru}}},
  \bibinfo{journal}{J. Phys.: Conf. Ser.} \textbf{\bibinfo{volume}{88}},
  \bibinfo{eid}{012044} (\bibinfo{year}{2007}).

\bibitem[{\citenamefont{Smith and Burch}(1959)}]{Smith1959}
\bibinfo{author}{\bibfnamefont{S.~J.} \bibnamefont{Smith}} \bibnamefont{and}
  \bibinfo{author}{\bibfnamefont{D.~S.} \bibnamefont{Burch}},
  \bibinfo{journal}{Phys. Rev. Lett.} \textbf{\bibinfo{volume}{2}},
  \bibinfo{pages}{165} (\bibinfo{year}{1959}).

\bibitem[{\citenamefont{{G{\'e}n{\'e}vriez} and {Urbain}}(2015)}]{gene15}
\bibinfo{author}{\bibfnamefont{M.}~\bibnamefont{{G{\'e}n{\'e}vriez}}}
  \bibnamefont{and} \bibinfo{author}{\bibfnamefont{X.}~\bibnamefont{{Urbain}}},
  \bibinfo{journal}{\pra} \textbf{\bibinfo{volume}{91}}, \bibinfo{eid}{033403}
  (\bibinfo{year}{2015}).

\bibitem[{\citenamefont{{Venuti} and {Decleva}}(1997)}]{venuti97}
\bibinfo{author}{\bibfnamefont{M.}~\bibnamefont{{Venuti}}} \bibnamefont{and}
  \bibinfo{author}{\bibfnamefont{P.}~\bibnamefont{{Decleva}}},
  \bibinfo{journal}{J. Phys. B: At. Mol. Opt. Phys.}
  \textbf{\bibinfo{volume}{30}}, \bibinfo{pages}{4839} (\bibinfo{year}{1997}).

\bibitem[{\citenamefont{{Feldmann}}(1977)}]{feldmann77}
\bibinfo{author}{\bibfnamefont{D.}~\bibnamefont{{Feldmann}}},
  \bibinfo{journal}{Chem. Phys. Lett.} \textbf{\bibinfo{volume}{47}},
  \bibinfo{pages}{338} (\bibinfo{year}{1977}).

\bibitem[{\citenamefont{Seman and Branscomb}(1962)}]{seman1962}
\bibinfo{author}{\bibfnamefont{M.~L.} \bibnamefont{Seman}} \bibnamefont{and}
  \bibinfo{author}{\bibfnamefont{L.~M.} \bibnamefont{Branscomb}},
  \bibinfo{journal}{Phys. Rev.} \textbf{\bibinfo{volume}{125}},
  \bibinfo{pages}{1602} (\bibinfo{year}{1962}).

\bibitem[{\citenamefont{{Branscomb} et~al.}(1965)\citenamefont{{Branscomb},
  {Smith}, and {Tisone}}}]{branscomb65}
\bibinfo{author}{\bibfnamefont{L.~M.} \bibnamefont{{Branscomb}}},
  \bibinfo{author}{\bibfnamefont{S.~J.} \bibnamefont{{Smith}}},
  \bibnamefont{and} \bibinfo{author}{\bibfnamefont{G.}~\bibnamefont{{Tisone}}},
  \bibinfo{journal}{J. Chem. Phys.} \textbf{\bibinfo{volume}{43}},
  \bibinfo{pages}{2906} (\bibinfo{year}{1965}).

\bibitem[{\citenamefont{{Hlavenka} et~al.}(2009)\citenamefont{{Hlavenka},
  {Otto}, {Trippel}, {Mikosch}, {Weidem{\"u}ller}, and {Wester}}}]{hlavenka09}
\bibinfo{author}{\bibfnamefont{P.}~\bibnamefont{{Hlavenka}}},
  \bibinfo{author}{\bibfnamefont{R.}~\bibnamefont{{Otto}}},
  \bibinfo{author}{\bibfnamefont{S.}~\bibnamefont{{Trippel}}},
  \bibinfo{author}{\bibfnamefont{J.}~\bibnamefont{{Mikosch}}},
  \bibinfo{author}{\bibfnamefont{M.}~\bibnamefont{{Weidem{\"u}ller}}},
  \bibnamefont{and} \bibinfo{author}{\bibfnamefont{R.}~\bibnamefont{{Wester}}},
  \bibinfo{journal}{J. Chem. Phys.} \textbf{\bibinfo{volume}{130}},
  \bibinfo{pages}{061105} (\bibinfo{year}{2009}).

\bibitem[{\citenamefont{{Dorsch} and {Daiminger}}(1999)}]{Dorsch1999}
\bibinfo{author}{\bibfnamefont{F.}~\bibnamefont{{Dorsch}}} \bibnamefont{and}
  \bibinfo{author}{\bibfnamefont{F.~X.} \bibnamefont{{Daiminger}}}, in
  \emph{\bibinfo{booktitle}{In-Plane Semiconductor Lasers III}}, edited by
  \bibinfo{editor}{\bibfnamefont{H.~K.} \bibnamefont{{Choi}}} \bibnamefont{and}
  \bibinfo{editor}{\bibfnamefont{P.~S.} \bibnamefont{{Zory}}}
  (\bibinfo{year}{1999}), vol. \bibinfo{volume}{3628} of
  \emph{\bibinfo{series}{Society of Photo-Optical Instrumentation Engineers
  (SPIE) Conference Series}}, pp. \bibinfo{pages}{56--63}.

\bibitem[{\citenamefont{Sun}(2015)}]{Sun2015}
\bibinfo{author}{\bibfnamefont{H.}~\bibnamefont{Sun}}, \emph{\bibinfo{title}{A
  practical guide to handling laser diode beams}}
  (\bibinfo{publisher}{SpringerBriefs in Physics, Dordrecht, Heidelberg, New
  York, London}, \bibinfo{year}{2015}).

\bibitem[{\citenamefont{Marshall and Stutz}(2011)}]{Marshall2011}
\bibinfo{author}{\bibfnamefont{G.~F.} \bibnamefont{Marshall}} \bibnamefont{and}
  \bibinfo{author}{\bibfnamefont{G.~E.} \bibnamefont{Stutz}},
  \emph{\bibinfo{title}{Handbook of optical and laser scanning}}
  (\bibinfo{publisher}{CRC Press}, \bibinfo{year}{2011}).

\bibitem[{zem()}]{zemax}
\bibinfo{note}{\protect http://www.zemax.com}.

\bibitem[{\citenamefont{Wang et~al.}(2009)\citenamefont{Wang, Yuan, Kang, Yang,
  Zhang, and Liu}}]{wang09}
\bibinfo{author}{\bibfnamefont{J.}~\bibnamefont{Wang}},
  \bibinfo{author}{\bibfnamefont{Z.}~\bibnamefont{Yuan}},
  \bibinfo{author}{\bibfnamefont{L.}~\bibnamefont{Kang}},
  \bibinfo{author}{\bibfnamefont{K.}~\bibnamefont{Yang}},
  \bibinfo{author}{\bibfnamefont{Y.}~\bibnamefont{Zhang}}, \bibnamefont{and}
  \bibinfo{author}{\bibfnamefont{X.}~\bibnamefont{Liu}}, in
  \emph{\bibinfo{booktitle}{Electronic Components and Technology Conference,
  2009. ECTC 2009. 59th}} (\bibinfo{year}{2009}), pp.
  \bibinfo{pages}{837--842}, ISSN \bibinfo{issn}{0569-5503}.

\bibitem[{\citenamefont{{Kreckel} et~al.}(2010)\citenamefont{{Kreckel},
  {Bruhns}, {{\v C}{\'{\i}}{\v z}ek}, {Glover}, {Miller}, {Urbain}, and
  {Savin}}}]{science2010}
\bibinfo{author}{\bibfnamefont{H.}~\bibnamefont{{Kreckel}}},
  \bibinfo{author}{\bibfnamefont{H.}~\bibnamefont{{Bruhns}}},
  \bibinfo{author}{\bibfnamefont{M.}~\bibnamefont{{{\v C}{\'{\i}}{\v z}ek}}},
  \bibinfo{author}{\bibfnamefont{S.~C.~O.} \bibnamefont{{Glover}}},
  \bibinfo{author}{\bibfnamefont{K.~A.} \bibnamefont{{Miller}}},
  \bibinfo{author}{\bibfnamefont{X.}~\bibnamefont{{Urbain}}}, \bibnamefont{and}
  \bibinfo{author}{\bibfnamefont{D.~W.} \bibnamefont{{Savin}}},
  \bibinfo{journal}{Science} \textbf{\bibinfo{volume}{329}},
  \bibinfo{pages}{69} (\bibinfo{year}{2010}).

\bibitem[{\citenamefont{Bruhns et~al.}(2010{\natexlab{a}})\citenamefont{Bruhns,
  Kreckel, Miller, Lestinsky, Seredyuk, Mitthumsiri, Schmitt, Schnell, Urbain,
  Rappaport et~al.}}]{bruhns2010}
\bibinfo{author}{\bibfnamefont{H.}~\bibnamefont{Bruhns}},
  \bibinfo{author}{\bibfnamefont{H.}~\bibnamefont{Kreckel}},
  \bibinfo{author}{\bibfnamefont{K.}~\bibnamefont{Miller}},
  \bibinfo{author}{\bibfnamefont{M.}~\bibnamefont{Lestinsky}},
  \bibinfo{author}{\bibfnamefont{B.}~\bibnamefont{Seredyuk}},
  \bibinfo{author}{\bibfnamefont{W.}~\bibnamefont{Mitthumsiri}},
  \bibinfo{author}{\bibfnamefont{B.~L.} \bibnamefont{Schmitt}},
  \bibinfo{author}{\bibfnamefont{M.}~\bibnamefont{Schnell}},
  \bibinfo{author}{\bibfnamefont{X.}~\bibnamefont{Urbain}},
  \bibinfo{author}{\bibfnamefont{M.~L.} \bibnamefont{Rappaport}},
  \bibnamefont{et~al.}, \bibinfo{journal}{Rev. Sci. Instrum.}
  \textbf{\bibinfo{volume}{81}}, \bibinfo{pages}{013112}
  (\bibinfo{year}{2010}{\natexlab{a}}).

\bibitem[{\citenamefont{Kreckel et~al.}(2010)\citenamefont{Kreckel, Bruhns,
  Miller, Wahlin, Davis, Hockh, and Savin}}]{Kreckel2010}
\bibinfo{author}{\bibfnamefont{H.}~\bibnamefont{Kreckel}},
  \bibinfo{author}{\bibfnamefont{H.}~\bibnamefont{Bruhns}},
  \bibinfo{author}{\bibfnamefont{K.~A.} \bibnamefont{Miller}},
  \bibinfo{author}{\bibfnamefont{E.}~\bibnamefont{Wahlin}},
  \bibinfo{author}{\bibfnamefont{A.}~\bibnamefont{Davis}},
  \bibinfo{author}{\bibfnamefont{S.}~\bibnamefont{Hockh}}, \bibnamefont{and}
  \bibinfo{author}{\bibfnamefont{D.~W.} \bibnamefont{Savin}},
  \bibinfo{journal}{Rev. Sci. Instrum.} \textbf{\bibinfo{volume}{81}},
  \bibinfo{pages}{063304} (\bibinfo{year}{2010}).

\bibitem[{sim()}]{simionweb}
\bibinfo{note}{\textsc{Simion 3D 8.0}, http://www.simion.com (2008)}.

\bibitem[{\citenamefont{Bruhns et~al.}(2010{\natexlab{b}})\citenamefont{Bruhns,
  Kreckel, Miller, Urbain, and Savin}}]{bruhns2010a}
\bibinfo{author}{\bibfnamefont{H.}~\bibnamefont{Bruhns}},
  \bibinfo{author}{\bibfnamefont{H.}~\bibnamefont{Kreckel}},
  \bibinfo{author}{\bibfnamefont{K.~A.} \bibnamefont{Miller}},
  \bibinfo{author}{\bibfnamefont{X.}~\bibnamefont{Urbain}}, \bibnamefont{and}
  \bibinfo{author}{\bibfnamefont{D.~W.} \bibnamefont{Savin}},
  \bibinfo{journal}{Phys. Rev. A} \textbf{\bibinfo{volume}{82}},
  \bibinfo{pages}{042708} (\bibinfo{year}{2010}{\natexlab{b}}).

\bibitem[{\citenamefont{{Miller} et~al.}(2011)\citenamefont{{Miller}, {Bruhns},
  {Eli{\'a}{\v s}ek}, {{\v C}{\'{\i}}{\v z}ek}, {Kreckel}, {Urbain}, and
  {Savin}}}]{miller2011}
\bibinfo{author}{\bibfnamefont{K.~A.} \bibnamefont{{Miller}}},
  \bibinfo{author}{\bibfnamefont{H.}~\bibnamefont{{Bruhns}}},
  \bibinfo{author}{\bibfnamefont{J.}~\bibnamefont{{Eli{\'a}{\v s}ek}}},
  \bibinfo{author}{\bibfnamefont{M.}~\bibnamefont{{{\v C}{\'{\i}}{\v z}ek}}},
  \bibinfo{author}{\bibfnamefont{H.}~\bibnamefont{{Kreckel}}},
  \bibinfo{author}{\bibfnamefont{X.}~\bibnamefont{{Urbain}}}, \bibnamefont{and}
  \bibinfo{author}{\bibfnamefont{D.~W.} \bibnamefont{{Savin}}},
  \bibinfo{journal}{\pra} \textbf{\bibinfo{volume}{84}}, \bibinfo{eid}{052709}
  (\bibinfo{year}{2011}).

\bibitem[{\citenamefont{{Miller} et~al.}(2012)\citenamefont{{Miller}, {Bruhns},
  {{\v C}{\'{\i}}{\v z}ek}, {Eli{\'a}{\v s}ek}, {Cabrera-Trujillo}, {Kreckel},
  {O'Connor}, {Urbain}, and {Savin}}}]{miller2012}
\bibinfo{author}{\bibfnamefont{K.~A.} \bibnamefont{{Miller}}},
  \bibinfo{author}{\bibfnamefont{H.}~\bibnamefont{{Bruhns}}},
  \bibinfo{author}{\bibfnamefont{M.}~\bibnamefont{{{\v C}{\'{\i}}{\v z}ek}}},
  \bibinfo{author}{\bibfnamefont{J.}~\bibnamefont{{Eli{\'a}{\v s}ek}}},
  \bibinfo{author}{\bibfnamefont{R.}~\bibnamefont{{Cabrera-Trujillo}}},
  \bibinfo{author}{\bibfnamefont{H.}~\bibnamefont{{Kreckel}}},
  \bibinfo{author}{\bibfnamefont{A.~P.} \bibnamefont{{O'Connor}}},
  \bibinfo{author}{\bibfnamefont{X.}~\bibnamefont{{Urbain}}}, \bibnamefont{and}
  \bibinfo{author}{\bibfnamefont{D.~W.} \bibnamefont{{Savin}}},
  \bibinfo{journal}{Phys. Rev. A.} \textbf{\bibinfo{volume}{86}},
  \bibinfo{eid}{032714} (\bibinfo{year}{2012}).

\bibitem[{\citenamefont{{von Hahn} et~al.}(2011)\citenamefont{{von Hahn},
  {Berg}, {Blaum}, {Crespo Lopez-Urrutia}, {Fellenberger}, {Froese}, {Grieser},
  {Krantz}, {K{\"u}hnel}, {Lange} et~al.}}]{vonhahn11}
\bibinfo{author}{\bibfnamefont{R.}~\bibnamefont{{von Hahn}}},
  \bibinfo{author}{\bibfnamefont{F.}~\bibnamefont{{Berg}}},
  \bibinfo{author}{\bibfnamefont{K.}~\bibnamefont{{Blaum}}},
  \bibinfo{author}{\bibfnamefont{J.~R.} \bibnamefont{{Crespo Lopez-Urrutia}}},
  \bibinfo{author}{\bibfnamefont{F.}~\bibnamefont{{Fellenberger}}},
  \bibinfo{author}{\bibfnamefont{M.}~\bibnamefont{{Froese}}},
  \bibinfo{author}{\bibfnamefont{M.}~\bibnamefont{{Grieser}}},
  \bibinfo{author}{\bibfnamefont{C.}~\bibnamefont{{Krantz}}},
  \bibinfo{author}{\bibfnamefont{K.-U.} \bibnamefont{{K{\"u}hnel}}},
  \bibinfo{author}{\bibfnamefont{M.}~\bibnamefont{{Lange}}},
  \bibnamefont{et~al.}, \bibinfo{journal}{Nucl. Instrum. Meth. Phys. Res. B}
  \textbf{\bibinfo{volume}{269}}, \bibinfo{pages}{2871} (\bibinfo{year}{2011}).

\end{thebibliography}

\end{document}